\documentclass[prb,
amsfonts,amssymb,floats,groupedaddress,superscriptaddress,twocolumn,aps,floatfix,longbibliography,10pt]{revtex4-2}

\usepackage{graphicx}
\usepackage{graphics}
\usepackage{amsmath}
\usepackage{amssymb}
\usepackage{amsfonts}
\usepackage{dsfont}
\usepackage{braket}
\usepackage{color}
\usepackage{braket,slashed}
\usepackage[mathscr]{euscript}
\definecolor{darkblue}{rgb}{0, 0, 0.8}
\usepackage[colorlinks=true, breaklinks=true, linkcolor=red, citecolor=blue, urlcolor=blue]{hyperref} 
\usepackage{hyperref}
\usepackage{subfigure}
\usepackage{xfrac}
\usepackage{bm}
\usepackage{kantlipsum}
\usepackage{enumitem}
\usepackage{tikz}
\usepackage{framed}
\usepackage{graphicx}
\usepackage{subfigure}
\usepackage{cleveref}
\usepackage{array}
\usepackage{nicefrac}

\usepackage{orcidlink}

\newcommand{\parL}[1]{\noindent\textbf{\textit{#1}}---}

\allowdisplaybreaks[1]

\newcommand{\code}[1]{\texttt{#1}}

\renewcommand{\dag}{^\dagger}

\renewcommand{\d}{\ensuremath{\mathrm{d}}}

\newcommand{\U}{\ensuremath{\mathrm{U}}}

\newcommand{\ulbaddress}{Center for Nonlinear Phenomena and Complex Systems, Universit\'e Libre de Bruxelles, CP 231, Campus Plaine, 1050 Brussels, Belgium}
\newcommand{\solvayaddress}{International Solvay Institutes, 1050 Brussels, Belgium}
\newcommand{\lkbaddress}{Laboratoire Kastler Brossel, Coll\`ege de France, CNRS, ENS-Universit\'e PSL, Sorbonne Universit\'e, 11 Place Marcelin Berthelot, 75005 Paris, France}

\usepackage{ulem}
\definecolor{mygreen}{rgb}{0,0.75,0}

\begin{document}

\title{Fractional Quantum Hall Wedding Cakes}

\author{Chloé Van Bastelaere\orcidlink{0009-0002-6164-6481}}
\email{chloe.van.bastelaere@ulb.be}
\author{Felix A. Palm\orcidlink{0000-0001-5774-5546}}
\author{Botao Wang\orcidlink{0000-0002-8220-2452}}
\affiliation{\ulbaddress}
\affiliation{\solvayaddress}

\author{Nathan Goldman\orcidlink{0000-0002-0757-7289}}
\affiliation{\ulbaddress}
\affiliation{\solvayaddress}
\affiliation{\lkbaddress}

\author{Laurens Vanderstraeten\orcidlink{0000-0002-3227-9822}}
\affiliation{\ulbaddress}

\date{\today}

\begin{abstract}

This work investigates the coexistence of distinct topologically ordered phases within a single setup. We demonstrate this concept through tensor network simulations of the Hofstadter-Bose-Hubbard model under a spatially modulated chemical potential. Focusing on cylindrical geometries, we realize regions exhibiting the Laughlin-1/2 phase and its particle-hole conjugate, and confirm their topological character via the local Středa's response and Laughlin's flux insertion protocol. Our approach offers a new pathway for experimentally and numerically charting entire phase diagrams within a single system, possibly eliminating the need for independent parameter scans.

\end{abstract}

\maketitle

\parL{Introduction}%
%
Quantum phases of matter are conceived as stable regions within the parameter space of quantum many-body systems at zero temperature. One exciting example is the fractional quantum Hall (FQH) effect: in a two-dimensional interacting electron gas under a strong magnetic field, many incompressible phases can be found with exotic physical properties described by the concept of topological order~\cite{moessner:21}. These phases are said to be robust because their topological signatures are stable against small variations of the system's macroscopic parameters. However, one expects that these phases will also be robust locally against sufficiently smooth \textit{spatial} variations of these parameters. This leads to the possibility of realizing different phases as spatially localized incompressible regions within a single system, separated by either sharp interfaces or extended compressible regions.

In the solid-state context, the coexistence of different types of topological order has gathered a lot of attention; a recent example is the debate on the origin of the half-integer thermal Hall conductance in the $\nu=5/2$ FQH plateau \cite{banerjee:18}, with the existence of ``puddles'' of distinct topological orders \cite{wang:18, mross:18, zhu:20, simon:20, lotric:25} as a possible explanation. Furthermore, this idea of realizing distinct phases in spatial regions constitutes an interesting experimental scenario in the context of quantum simulation, where controlling and probing quantum many-body systems locally has seen tremendous progress in recent years~\cite{gross:17}. A standard example is the realization of ``wedding cake''-like structures in bosonic systems, with Mott insulators at different fillings separated by superfluid regions~\cite{jaksch:98, folling:06, sherson:10}. On the other hand, the realization of extended topological quantum phases has been made feasible by the advent of artificial gauge fields for neutral atoms~\cite{aidelsburger:13,miyake:13}, even in the strongly interacting regime~\cite{tai:17}. This has recently culminated in the first experimental realizations of FQH states of cold atoms in optical lattices~\cite{leonard:23} and rotating traps~\cite{Lunt2024}, as well as in photonic systems~\cite{Wang2024}. 
\begin{figure}[b]
\includegraphics[scale=0.85]{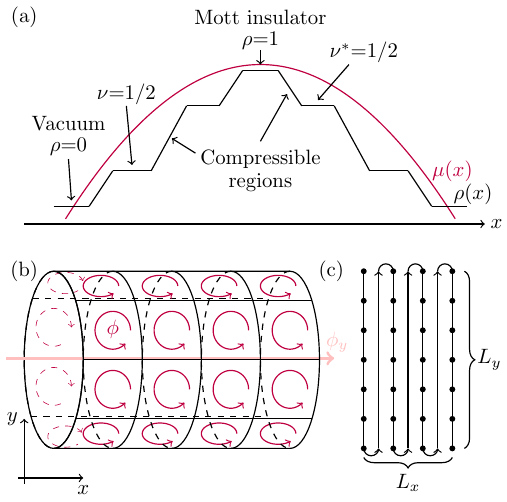}
\caption{\textbf{(a)} Sketch of a wedding cake-like structure with FCI regions in a system with a trapping potential. \textbf{(b)} Illustration of the Hofstadter-Bose-Hubbard model on a cylinder and, \textbf{(c)} a $L_x\times L_y$ square lattice with MPS snaking.  
}
\label{fig:mps}
\end{figure}

In this work, we numerically explore the possibility of realizing multiple incompressible FQH phases in a single system by adding a confining potential, see Fig.~\ref{fig:mps}(a). We investigate this phenomenon in the Hofstadter-Bose-Hubbard model, which describes interacting bosons in a magnetic field on a two-dimensional lattice~\cite{hofstadter:76, harper:55, azbel:64}. This model is known to host a large number of lattice analogues of FQH phases, known as fractional Chern insulators (FCI)~\cite{sorensen:05, moller:09, regnault:11, sun:11, neupert:11, tang:11, sheng:11, scaffidi:12, bergholtz:13, parameswaran:13}. Relying on tensor network~\cite{schollwock:11, cirac:21} simulations, we characterize the different phases by computing the entanglement entropy and extracting the many-body Chern number through a local St\v{r}eda response and the flux insertion protocol.

\parL{Model and methods}%
%
We study the bosonic Hofstadter-Hubbard model with nearest-neighbor hopping amplitude $t$ and on-site interaction strength $U$ on a finite $L=L_x\times L_y$ square lattice.
We impose periodic boundary conditions in the $y$-direction and open boundary conditions in the $x$-direction in order to realize a cylinder geometry, see Fig.~\ref{fig:mps}(b).
The system is subject to an external magnetic field, which results in a magnetic flux piercing through each plaquette of the lattice.
The flux per plaquette is defined as $\phi = 2\pi \alpha$, where $\alpha$ is a rational magnetic flux density.
For future purposes, we also introduce a magnetic flux $\phi_y$ inserted across the cylinder, as shown in Fig.~\ref{fig:mps}(b).
We consider a chemical potential $\mu(x)$ that is inhomogeneous in the $x$-direction and invariant in the $y$-direction.
In the Landau gauge, the Hamiltonian is given by, with $\tilde{L}_x=\lfloor L_x/2\rfloor,$
\begin{equation}
	\begin{aligned}
		H = &- t\sum_{x=1-\tilde{L}_x}^{\tilde{L}_x}\sum_{y=1}^{L_y} \left(e^{i\left(\phi x+\phi_y/L_y\right)} a_{x,y}\dag a_{x,y+1}+ \mathrm{h.c.}\right) \\
        &- t\sum_{x=1-\tilde{L}_x}^{\tilde{L}_x-1}\sum_{y=1}^{L_y}\left( a_{x,y}\dag a_{x+1,y} + \mathrm{h.c.}\right)\\
			&+\sum_{x=1-\tilde{L}_x}^{\tilde{L}_x}\sum_{y=1}^{L_y}\left(\frac{U}{2} n_{x,y}(n_{x,y}-1) - \mu(x) n_{x,y}\right), \\
	\end{aligned}
\end{equation}
where $a_{x,y}^{(\dagger)}$ are the bosonic annihilation (creation) operators acting at site $(x,y)$ and $n_{x,y}=a_{x,y}\dag a_{x,y}$ is the bosonic number operator.
The characteristic length scale in the system is the magnetic length, defined as \mbox{$\ell_B=(2\pi\alpha)^{-1/2}$}.

In our simulations, we exclusively consider the hard-core bosons limit, $U/t\to\infty$, which we implement by imposing the maximum number of bosons per site to one. In this limit, the system is invariant under a particle-hole (PH), combined with either time-reversal symmetry or reflection symmetry. The latter combination is preserved with a varying chemical potential under the condition $\mu(-x)=-\mu(x)$, which implies $\rho(-x)=1-\rho(x)$. 

The realization of FCI phases in this model, or closely related ones, has been the subject of many numerical investigations using different approaches such as exact diagonalization on small systems~\cite{sorensen:05, hafezi:07, moller:15, andrews:18, repellin:19, repellin:20,mantas:18}, matrix product states (MPS) on finite~\cite{rosson:19, palm:21} and infinite cylinders~\cite{grushin:15, boesl:22, motruk:15, he:17, schoonderwoerd:19, motruk:17-b, dong:18, andrews:21}, projected entangled pair states on the infinite plane~\cite{weerda:24}, or tree tensor networks~\cite{gerster:17, macaluso:20}. In this work, we use MPS methods, as these are well suited to study the thin-cylinder geometry. We consider a path snaking around a cylinder that maps our two-dimensional system onto a one-dimensional one, see Fig.~\ref{fig:mps}(c). We then optimize variational MPS ground-state approximations on this effective 1-D model. In order to study the bulk properties of a given phase, we use infinite MPS simulations using either the iDMRG~\cite{mcculloch:08} or the VUMPS algorithm~\cite{zaunerstauber:18, vanderstraeten:19} to variationally obtain the ground state. On the finite cylinder geometry, we use the density matrix renormalization group (DMRG) algorithm~\cite{white:92,schollwock:11} to variationally optimize a finite MPS. The MPS bond dimension $D$ serves as a control parameter in our simulations. The maximal bond dimension used here is $D_{\max}=1000$. Similar results were obtained with lower maximal bond dimension, showing the convergence of the simulations. In both the finite and infinite cases, the numerical cost scales exponentially with the cylinder circumference. 

In the hard-core limit, the bulk phase diagram of the model as a function of chemical potential has been largely established \cite{boesl:22}. The phases are defined by the filling fraction $\nu=\rho/\alpha$, where $\rho=N_b/L$ is the boson density and $N_b$ is the number of bosons. To describe the PH conjugate states, we introduce a hole-filling fraction $\nu^*=(1-\rho)/\alpha$. The Laughlin state~\cite{laughlin:83} shows up as a sizeable incompressible region (i.e., $\partial\rho/\partial\mu=0)$ at a magnetic filling factor $\nu=1/2$, as well as its PH conjugate at $\nu^*=1/2$. We estimate the size $\Delta_\mu$ of this region by computing the filling fraction as a function of the chemical potential $\mu$ by grand-canonical ground state searches. We obtain a size around $\Delta_{\mu}\sim 0.3 t$, which agrees with the results found in previous studies~\cite{sorensen:05}. By PH symmetry, the region at $\nu^*=1/2$ has the same size. For later reference, we calculate the correlation length $\xi$ in the $\nu=1/2$ and $\nu^*=1/2$ states using infinite MPS simulations and obtain $\xi/\ell_B\sim2$. We note that there exist other plateaux for different values of $\nu$, corresponding to different FCI states. Their respective sizes are typically smaller than the size of the Laughlin-1/2 phase, which makes their realization tedious~\cite{palm:21,boesl:22}.

In previous works, it has been observed that FCI plateaux exhibit spurious density oscillations on the elongated cylinder geometry~\cite{palm:21}, hindering our goal of revealing spatially localized density plateaux. Choosing $\alpha=2/L_y$, we can avoid these charge density wave instabilities~\cite{palm:25}. With this choice, we find that $L_y=7$ is large enough to stabilize large $\nu=1/2$ and $\nu^*=1/2$ plateaux while still enabling efficient MPS simulations. For the rest of this work, we work exclusively with these parameters.

\parL{Density profiles}%
%
\begin{figure}[t]
    \centering
    \includegraphics{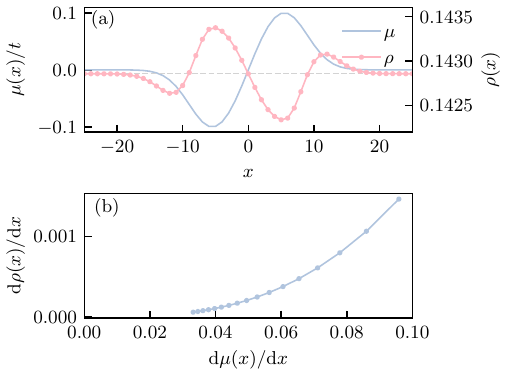}    
    \caption{\textbf{(a)} The variation of the density $\rho(x)$ as the chemical potential $\mu(x)$ changes in a local region, embedded within a uniform FCI state. The dashed line represents the density $\rho_0$. $\textbf{(b)}$ The slope of the density as a function of the variation of the chemical potential in the center of the local region (at $x=0$), obtained from varying the profile $\mu(x)$. The solid line serves as a guide for the eye.
    }
    \label{fig:lda}
\end{figure}
In this section, we study the appearance of incompressible regions in the system by computing the local boson density,
\begin{equation}
	\rho(x) = \frac{1}{L_y} \sum_{y=1}^{L_y} \langle n_{x,y} \rangle, \quad x=1,\dots,L_x,
\end{equation}
across a $L_x\times L_y=90\times 7$ cylinder with a varying chemical potential $\mu(x)$.

In order to realize stable FCI regions in the presence of a varying chemical potential, we rely on the local density approximation~\cite{cooper:05}:
we assume that the density remains constant as long as the varying chemical potential $\mu(x)$ lies within the stability region of the uniform phase.
Let us first investigate to what extent this approximation is justified on the $\nu=1/2$ plateau. Starting from the uniform $\nu=1/2$ state on an infinite cylinder with boson density $\rho_0$, we vary the chemical potential $\mu(x)$ in a localized region and compute the resulting variation of the local density. In order to approximate the ground state in this local region, we optimize a non-uniform ``segment MPS'' embedded in an infinite MPS~\cite{Phien2012} with $\U(1)$ symmetry. 
To preserve the total number of particles in the system, we use an odd chemical potential $\mu(x)$, so that the resulting density deviation $\rho(x)$ from $\rho_0$ is also odd. We make sure that the chemical potential stays within the bulk stability region of the FCI phase, i.e., $|\mu(x)|<\Delta_\mu$. In Fig.~\ref{fig:lda}(a), we observe that the density variations stay extremely small, even for quite large variations of $\mu(x)$. We repeat this numerical experiment with different profiles $\mu(x)$, and systematically track the response in the density. In Fig.~\ref{fig:lda}(b), we show the slope of the density response as a function of the slope in the chemical potential in the middle of the region. From this figure, we can thus see that a varying chemical potential leads to extremely small density variations, which validates the use of the local density approximation on an FCI plateau.

Given that the local density approximation holds within a given incompressible region, we need to ensure that these regions are wide enough so that we can resolve the topological signatures. For that purpose, we demand the size of the region to be larger than the correlation length. This translates into the requirement:
\begin{equation}\label{eq:chargegap}
	\frac{\d \mu(x)}{\d x} < \frac{\Delta_{\mu}}{\xi}.
\end{equation}
In this way, we can be sure that the edges on either side of the plateau cannot couple, and we expect to see the local signatures of the FCI phase.

\begin{figure}[t!]
    \includegraphics{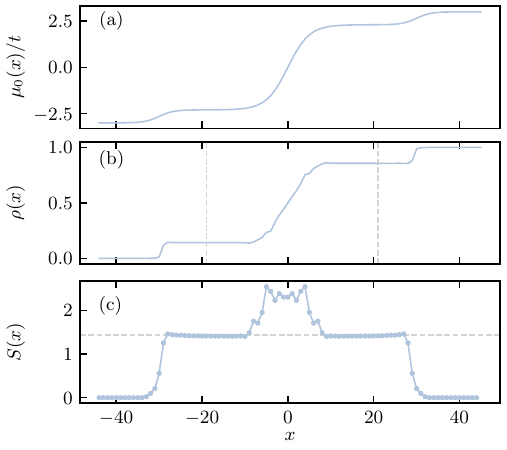}
    \caption{\textbf{(a)} The chemical potential $\mu_0(x)$, and
        \textbf{(b)} the density profile in a $90\times7$ cylinder with $\alpha=2/7$. 
        The dashed lines represent the tripartition we make for the charge pumping experiment.
        \textbf{(c)} The entanglement entropy $S(x)$.
        The dashed line is the value of the entanglement entropy for a Laughlin-1/2 state in a uniform MPS simulation.
    }
    \label{fig:densityadiabatic0}
\end{figure}
We start by considering a smooth chemical potential $\mu_0(x)$ that is defined by a sum of three hyperbolic tangent functions, as represented in Fig.~\ref{fig:densityadiabatic0}(a).
The chemical potential $\mu_0(x)$ takes values in the range $[-3t,3t]$ in which different FCI phases with filling fraction $0\leq \nu \leq 1/\alpha$ can be found.
This chemical potential realizes regions of constant density, as it contains different plateaux that we position to target the Laughlin-1/2 and its PH conjugate states.

In Fig.~\ref{fig:densityadiabatic0}(b), we compute the density profile along the cylinder with $\alpha=2/7$. First, we observe the realization of two trivial phases at $\rho=0$ and $\rho=1$ corresponding to the vacuum state and a Mott insulator state.
We also obtain incompressible plateaux at fractional filling fractions $\nu=1/2$ and $\nu^*=1/2$ ($\nu=3$), which correspond to the Laughlin-1/2 phase and its PH conjugate.
In between the two FCI plateaux, we find an extended compressible region in which the density varies smoothly as a function of $\mu(x)$.

In order to further confirm that the fractional density plateaux correspond to the Laughlin state and its PH conjugate, we compute the entanglement entropy $S(x)$ corresponding to a bipartition of the cylinder at a given location $x$.
In Fig.~\ref{fig:densityadiabatic0}(c), we see that the entropy is zero for the two trivial states, as expected.
For the two FCI phases we obtain a plateau of constant entanglement entropy. Its value coincides with the one we obtain from simulations of the Laughlin-1/2 state and its PH conjugate with a constant chemical potential.

\begin{figure}
    \includegraphics{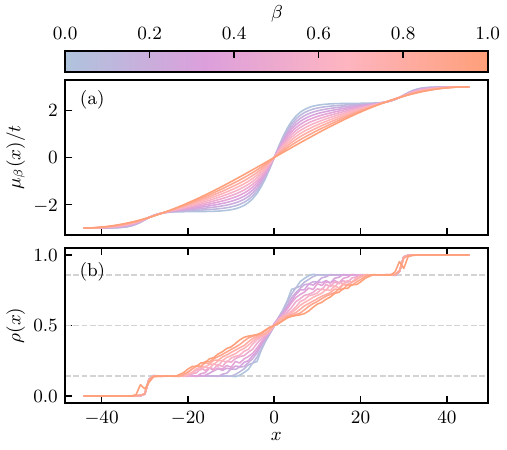}
    \caption{\textbf{(a)} The adiabatic change of the chemical potential $\mu_{\beta}(x)$ for different values of $\beta$.
        \textbf{(b)} The density profiles in a $90\times7$ cylinder with $\alpha=2/7$ for different values of $\beta$. The dashed lines show $\rho=1/7,1/2,6/7$ as a guide for the eye.
    }
    \label{fig:densityadiabatic}
\end{figure}

This first trivial choice of chemical potential allows for the clear observation of stable FCIs and is nowadays realizable in cold-atom experiments by using digital micromirror devices (DMDs)~\cite{Zupancic2016} or spatial light modulators (SLMs)~\cite{Gaunt2012}.
However, we should also consider chemical potential profiles without constant regions, for which the occurrence of density plateaux is not as straightforward.
In an experimental setup, the most natural choice is a quadratic chemical potential that forms a harmonic trap. As we aim to preserve the PH symmetry in the system, it is, however, more convenient to work with a sine squared chemical potential that satisfies Eq.~\eqref{eq:chargegap} for all $x$,
\begin{equation}\label{eq:sine}
	\mu_1(x) = 6 \sin^2\left(\frac{\pi}{2L_x}\left(x+\frac{L_x}{2}\right)\right)-3,
\end{equation}
where the additive and multiplicative constants are chosen to reach the desired states.
In order to show the evolution of the FCI plateaux, we smoothly deform the chemical potential starting from $\mu_0(x)$ into $\mu_1(x)$ using
\begin{equation}\label{eq:adiabaticpotential}
	\mu_{\beta}(x) = (1-\beta)\mu_0(x) + \beta \mu_1(x),
\end{equation}
with $\beta$ varying between $0$ and $1$, as represented in Fig.~\ref{fig:densityadiabatic}(a).
The density profiles of this adiabatic change are represented in Fig.~\ref{fig:densityadiabatic}(b).
We observe that the size of the FCI regions obtained with $\beta=0$ slowly reduces to a few lattice sites, while the compressible one increases as $\beta\to1$.
Crucially, we observe that the plateaux persist up to $\beta=1$ and thus, they should have the same topological nature than the ones observed with $\beta=0$.
We note that the PH symmetry around $\rho=1/2$ is respected for every value of $\beta$.
%

\parL{Topological signatures}%
%
We now perform different numerical simulations to determine the topological signatures of the plateaux observed in Fig.~\ref{fig:densityadiabatic} to confirm the Laughlin-1/2 nature of the phases~\cite{laughlin:83}. In these investigations, we always use the chemical potential $\mu_0(x)$, for which the plateaux have the largest size. This makes the computation of their topological signatures clearer. However, the numerical extraction of the signatures remains challenging. Indeed, one observation from our investigations is the presence of many low-lying states due to the presence of the large compressible region. This phase typically requires much higher bond dimension to be well captured by tensor network simulations. This results in many instabilities and difficulties in the following numerical experiments to obtain clear topological signatures for the incompressible regions. Moreover, we note that the MPS approximation necessarily breaks the $\U(1)$ symmetry of the system in the compressible region~\cite{vanderstraeten:19b}, so that the total charge is not conserved in our simulations. 

\begin{figure}
    \includegraphics{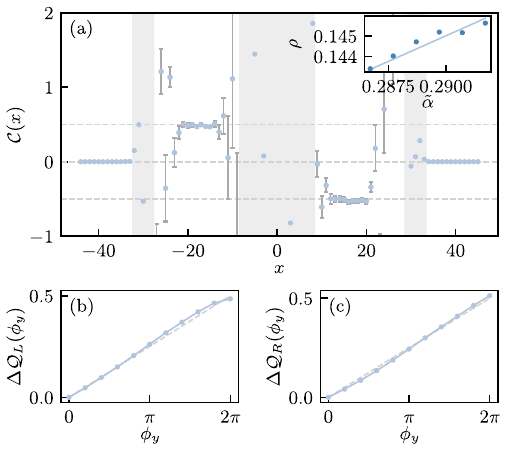}
    \caption{
        \textbf{(a)} The many-body Chern number obtained along the cylinder using Středa's response.
        The blue dots are the numbers obtained from our simulations.
        The dashed lines show the $\mathcal{C}=0,\pm 1/2$ values as a guide for the eye.
        We show the error bars from the fit in the FCI regions. The shaded regions correspond to the compressible phases where Středa's response is not meaningful.
        The inset shows the evolution of the density as a function of $\tilde{\alpha}=\alpha+\delta\alpha$ in the bulk of the Laughlin-1/2 region at $x=23$.
        The solid line is the linear fit we use to extract the Chern number.
        \textbf{(b)} The charge evolution as a function of the inserted flux $\phi_y$ for the left of the $\nu=1/2$ region, $\Delta \mathcal{Q}_L(\phi_y)$, and \textbf{(c)} the right of the $\nu^*=1/2$ region, $\Delta \mathcal{Q}_R(\phi_y)$. For each step of the adiabatic insertion, we perform DMRG ground state searches with a maximal bond dimension of $D_{\mathrm{max}}=200$.}
    \label{fig:streda_charge}
\end{figure}

One defining property of a topological phase is its many-body Chern number $\mathcal{C}$~\cite{tao:86,niu:85}. Although this is typically thought of as a global property of a system, we can access a spatially resolved many-body Chern number through Středa's formula~\cite{streda:82,streda:82-1},
\begin{equation}\label{eq:streda}
	\mathcal{C}(x) =  \frac{\partial \rho(x)}{\partial \alpha}.
\end{equation}
As the St\v{r}eda response is a local topological marker \cite{repellin:20,umucalilar:08,bianco:13}, we compute it for each $x$ across the cylinder.
We first compute multiple density profiles with shifted magnetic flux $\tilde{\alpha}=\alpha+\delta\alpha$ and use a linear fit to extract the Chern number for each $x$ of the cylinder.
As noted previously, the presence of the large compressible phase results in many instabilities, which requires us to reject some ill-converged simulations.
In Fig.~\ref{fig:streda_charge}(a), we show $\mathcal{C}(x)$ obtained along the $x$-axis and the linear fit used for a representative point in the bulk of the $\nu=1/2$ plateau in the inset.
We observe clear regions with $\mathcal{C}=0$ for the trivial states, as expected.
Furthermore, we obtain regions with Chern numbers $\mathcal{C}=1/2$ (resp. $\mathcal{C}=-1/2$) in the region where the $\nu=1/2$ (resp. $\nu^*=1/2$) phase is localized.
This agrees with the expectation of a Laughlin-1/2 and its PH conjugate phases.
However, as Středa's formula probes a bulk property, it is only expected to give a quantized value away from the edges of a plateau.
This results in the regions defined by $\mathcal{C}=\pm 1/2$ in Fig.~\ref{fig:streda_charge}(a) being smaller than their corresponding density plateaux of Fig.~\ref{fig:densityadiabatic0}(b).

A second way of accessing the topological properties of the plateaux is by conducting a charge pumping experiment~\cite{zaletel:14,grushin:15,wang:22,laughlin:81}.
To this end, we adiabatically insert a flux $\phi_y$ across the cylinder, see Fig.~\ref{fig:mps}.
We compute the ground state of the system with $\beta=0$ at each step of the adiabatic insertion.
We make a tripartition of our system by cutting in the middle of each FCI region as represented in Fig.~\ref{fig:densityadiabatic0}.
We then keep track of the charge evolution to the left of the first cut, $\Delta \mathcal{Q}_L(\phi_y)$, and to the right of the second cut, $\Delta \mathcal{Q}_R(\phi_y)$.
The Hall conductivity of each plateau is obtained after the insertion of $2\pi$ flux by
\begin{equation}
    \sigma_H = \frac{e^2}{h}\left(\mathcal{Q}_{L(R)}(\phi_y=2\pi)-\mathcal{Q}_{L(R)}(\phi_y=0)\right),
\end{equation}
where $\mathcal{Q}_{L(R)}(\phi_y)$ denotes the charge to the left (right) of the cut in the state $\nu=1/2$ ($\nu^*=1/2$) at flux $\phi_y$~\cite{zaletel:14,grushin:15}.
In Fig.~\ref{fig:streda_charge}(b), after the insertion of $2\pi$, we find results close to $\Delta \mathcal{Q}_L(2\pi) = +1/2$.
These results are consistent with the Laughlin-1/2 nature of the \mbox{$\nu=1/2$} phase.
As the PH conjugate state $\nu^*=1/2$ has the opposite chirality, the charge is now pumped in the other direction, so we find the same accumulation of charge on the right of the plateau, $\Delta \mathcal{Q}_R(2\pi)=+1/2$.

Moreover, as the total number of bosons in our system is not conserved, the compressible region does not follow this charge transfer pattern.
We have conducted the same experiment on a system without any compressible region (e.g., by considering a step-function-like chemical potential) and we find that the total number of bosons is conserved throughout the flux insertion protocol, obtaining $\Delta \mathcal{Q}_M=-1$ in the middle region (not shown).

\parL{Outlook}%
In this work, we have shown that a smoothly varying chemical potential can lead to the coexistence of fractional incompressible plateaux surrounded by compressible regions, and that the topological signatures of the incompressible regions can be probed all at once via local density measurements routinely performed in quantum gas microscopes~\cite{leonard:23}.
Our setup therefore constitutes an instructive example of how to realize a rich phase diagram within a single many-body state.
We expect that it can be extended to host even more incompressible regions with possibly more intricate types of topological order. It is known that FCI phases beyond the Laughlin-1/2 phase and its PH conjugate can be stabilized in this model~\cite{boesl:22, palm:21}.
From our work, however, it is clear that realizing more phases simultaneously would require much larger system sizes, possibly beyond the reach of the current numerical and experimental approaches.

In order to minimize edge effects while keeping the system size manageable for our numerical simulations, we have worked with a cylindrical geometry with a circumference of $L_y=7$.
In an experimental context, an open strip geometry or a radially varying chemical potential is likely to be more realistic.
For the former, the occurrence of gapless edge modes might decrease the stability of the FCI plateaux, and a proper engineering of the confining potential in the $y$ direction would be needed.
We expect no dramatic changes in a radially varying setup, although the numerical simulations would become a lot more demanding, necessitating other tensor network approaches such as tree tensor networks~\cite{gerster:17} or finite projected entangled-pair states~\cite{verstraete:04}.

Our setup calls for further investigations into the interplay and coexistence of different types of (possibly topological) incompressible and (possibly superfluid) compressible regions. One type of question concerns the universal features of the interfaces between different regions. It would be particularly appealing to explore fermionic generalizations of our scheme, potentially leading to superconductor–FQH interfaces that host parafermions~\cite{PhysRevX.4.011036}. Finally, a more practical question is whether this coexistence of phases can be leveraged to develop more efficient preparation schemes. For instance, in the spirit of Ref.~\onlinecite{yang2020cooling}, one can envisage superfluid layers acting as reservoirs that effectively reduce the entropy within the (potentially numerous) FQH regions.

\parL{Acknowledgements}%
%
We would like to thank Julian Boesl and Frank Pollmann for discussions. This research was financially supported by the FRS-FNRS (Belgium), the ERC Grant LATIS, the EOS project CHEQS, and the Fondation ULB. Computational resources have been provided by the Consortium des Equipements de Calcul Intensif (CECI), funded by the F.R.S.-FNRS under Grant No. 2.5020.11 and by the Walloon Region. Tensor network simulations are implemented using ITensor~\cite{itensors1}, MPSKit~\cite{mpskit}, and TensorTrack~\cite{tensortrack}.

\bibliography{bibliography.bib}

\end{document}